\documentclass{2314aa}
\usepackage{graphics}
\usepackage{natbib}

\begin{document}

\title{A Serendipitous {\it XMM-Newton} Observation of the
Intermediate Polar WX Pyx
  \thanks{Research supported by contract number NAS8-39073 to SAO.}}

\author{Eric M. Schlegel}

\institute{High Energy Astrophysics Division, Smithsonian Astrophysical
Observatory, Cambridge, MA 02138; email: eschlegel@cfa.harvard.edu}

\date{Received 2004 November 4 / Accepted 2004 December 8}

\abstract{ We briefly describe a serendipitous observation of the
little-studied intermediate polar WX Pyx using {\it XMM-Newton}.  The
X-ray spin period is 1557.3 sec, confirming the optical period
published in 1996.  An orbital period of $\sim$5.54 hr is inferred
from the separation of the spin-orbit sidelobe components.  The soft
and hard band spin-folded light curves are nearly sinusoidal in shape.
The best-fit spectrum is consistent with a bremsstrahlung temperature
of $\sim$18 keV.  An upper limit of $\sim$300 eV is
assigned to the presence of Fe line emission.  WX Pyx lies near TX and
TV Col in the P$_{\rm spin}$-P$_{\rm orb}$ plane.
}

\titlerunning{XMM Observation of the IP WX Pyx}

\maketitle

\section{Introduction}

WX Pyx, originally designated 1E 0830.9-2238, was uncovered in a
survey of the Galactic plane using {\it Einstein} \citep{Hertz1984}
and described in more detail by \cite{Hertz1990}.  Those authors
identified the probable counterpart by searching for objects with the
highest UV excess.  An optical spectrum of 1E 0830 showed emission
lines typical of a cataclysmic variable (= CV), including H${\beta}$,
H${\gamma}$, and He II.  The large value of the He~II/H${\beta}$
emission ratio suggested a magnetic CV although that ratio is not a
unique indicator \citep{Szkody1990}.  1E 0830 acquired its new name on
the 71st Namelist of Variable Stars \citep{Kazarovets1993}.

Remarkably, there has been no detailed follow-up of this CV in the
X-ray band and little in any waveband.  \cite{odonoghue1996} supplied
the only detailed study to date using optical photometry.  They
uncovered a stable period of 26 min they assumed was the spin period
and indications of an orbital period of very roughly 6 hours and
raised the distinct possibility that the counterpart was an
intermediate polar (IP).  It was included in three surveys for radio
emission \citep{Pavelin1994, Beasley1994, Wendker1995} and a survey of
infrared emission from the 2MASS survey \citep{Hoard2002}.  As with
most CVs, WX Pyx was not detected in the radio but was seen in the IR.
\cite{Buckley2000} included it on a list of `to be observed' objects
in a summary of the power spectra of IPs.  To date, the spin and
orbital periods remain unconfirmed.

\begin{figure*}[t!]
\centering
\caption{Soft and hard light curves and hardness ratios of WX Pyx for
May 2003 PN data.}
\label{hrcurve}
\scalebox{0.5}{\rotatebox{-90}{\includegraphics{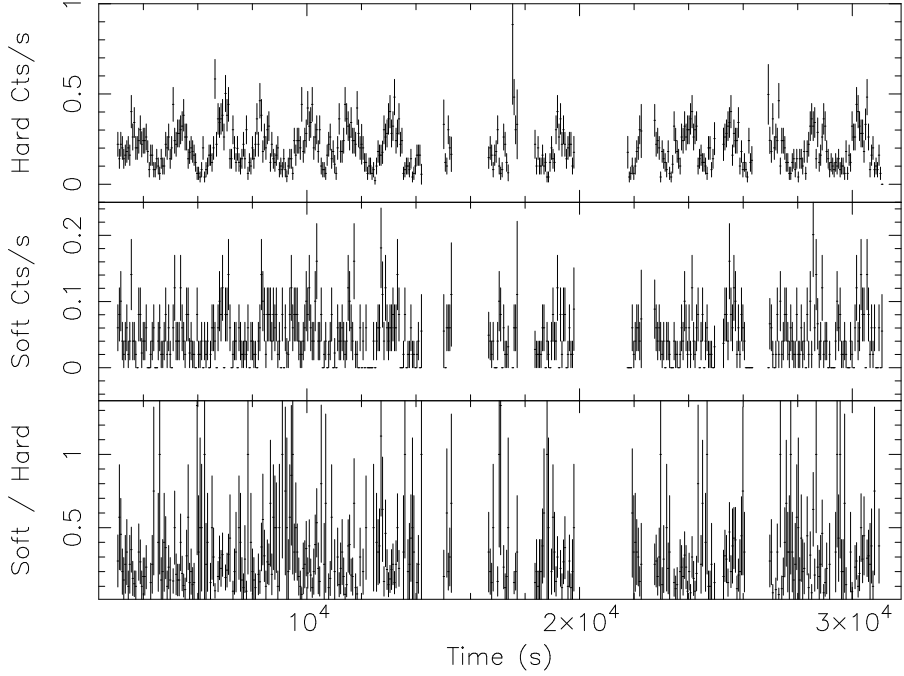}}}
\end{figure*}

Intermediate polars (= IP) are a subclass of magnetic CVs in which the
magnetic moment is insufficiently strong to dominant the dynamics of
the accretion flow completely.  As a result, the spin period of the
accreting white dwarf is not magnetically locked to the orbital period
as in the polars (eg, \citealt{Warner1995}).  The defining
characteristic of an IP is the presence of a strong pulse in the X-ray
band marking the spin of the white dwarf.  IPs generally show hard
spectra with a bremsstrahlung kT in the 10-30 keV range and strong Fe
K line emission.

We describe a serendipitous observation of WX Pyx by {\it XMM-Newton}.

\begin{table*}[t!]
\centering
\caption{{\it XMM-Newton} Observations of WX Pyx}
\label{obslist}
\begin{tabular}{r r l r r r l}
\hline \hline
     &       &      & Original & Filtered & Source & Detectors \cr
 No. & Obsid & Date & ExpT & ExpT & Count rate & Used \cr \hline
  1 & 0149160101 & 2003 Apr 23 & 45255 & 9478 & 0.04 & PN only \cr
  2 & 0149160201 & 2003 May 20 & 29919 & 25189 & 0.24 & PN \cr 
    &  $\cdots$  & $\cdots$   & $\cdots$ & 27250 & 0.07 & MOS-1 \cr
    &  $\cdots$  & $\cdots$   & $\cdots$ & 27299 & 0.05 & MOS-2 \cr \hline
\end{tabular}
\end{table*}

\section{Observation}

WX Pyx was observed for $\sim$45 ksec during a pointed observation of
NGC 2613, PI = T. Chaves, on 2003 Apr 23 (obsid 0149160101).  WX Pyx
lies $\sim$10.5$'$ offaxis to the northwest.  The large off-axis angle
ensures that only data from the EPIC-PN and -MOS detectors includes WX
Pyx; the source lies outside the fields-of-view of the Optical Monitor
and Reflection Gratings.  The April observation was hammered by soft
X-ray flares; at best, $\sim$20\% of the observation may be salvaged.
A second observation of NGC 2613 occurred on 2003 May 20 for 29299 sec
(obsid 0149160201).  This observation contains considerably fewer
flares, losing $<$30\% of the data to flaring activity.  For both
observations, the EPIC-PN used the thin filter.  Table~\ref{obslist}
summarizes the available data.  Figure~\ref{hrcurve} shows the
filtered light curves for the May EPIC-PN data in the soft (0.5-2 keV)
and hard (2-10 keV) bands as well as the hardness ratio (hard / soft).

\begin{figure}
\centering
\caption{Power spectrum for May EPIC-PN observation of WX Pyx.  The
peaks near 6.4$\times$10$^{-4}$ are identified as the
${\omega}-{\Omega}$, ${\omega}$, and ${\omega}+{\Omega}$ frequency
components, respectively.  The dashed line is the approximate
3${\sigma}$ detection threshold.}
\scalebox{0.3}{\rotatebox{-90}{\includegraphics{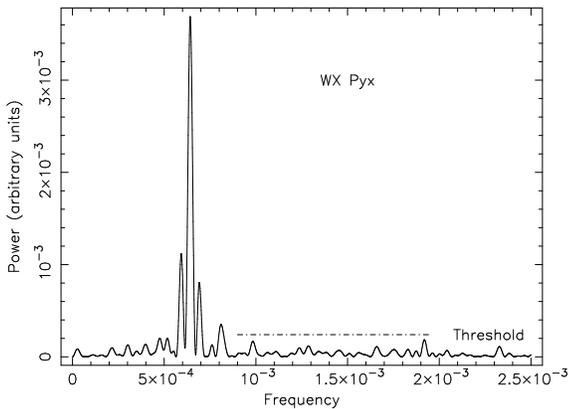}}}
\label{powspec}
\end{figure}

\section{Light Curve and Power Spectrum}

From an extracted light curve, times of flaring activity were
identified and subsequently filtered from the data sets.  Even in
perfect data sets, the count rate in the MOS detectors is perhaps a
third of the EPIC-PN detector.  Given the flaring in the April
observation, only the first 9 ksec could be used; combining the lower
count rate for the MOS detectors and the short data span, we used only
the EPIC-pn data.  For the May observation, data from the PN and MOS
detectors were used, suitably filtered to eliminate the times of
flares (Table~\ref{obslist}).

Applying an FFT to the full-band light curve yielded the power
spectrum shown in Figure~\ref{powspec}.  Three gaussians plus a
constant were fit to the data to determine the centroids of the power
spectrum components.  The peak is identified as the spin period
${\omega}$ of the IP; the two lower peaks are the sidelobes of
${\omega}-{\Omega}$ and ${\omega}+{\Omega}$.  The fitted spin
frequency is 6.42146$\pm$0.00075${\times}$10$^{-4}$ s$^{-1}$,
corresponding to a spin period of 1557.3$\pm$0.3 sec where the error
is conservatively adopted as the 90\% confidence range on the center
of the gaussian.

\begin{figure}
\centering
\caption{Normalized X-ray light curves in the (top) 0.5-2 and (bottom)
2-10 keV bands and folded at the measured spin period.  The
normalization factors are soft: 0.0459 cts s$^{-1}$ and hard: 0.2013
cts s$^{-1}$.}
\scalebox{0.3}{\rotatebox{-90}{\includegraphics{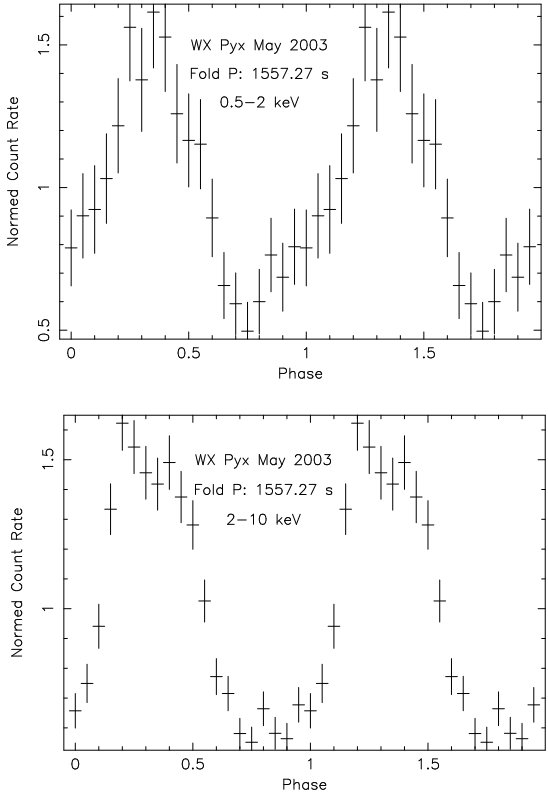}}}
\label{foldLC}
\end{figure}

\begin{figure}
\centering
\scalebox{0.3}{\rotatebox{-90}{\includegraphics{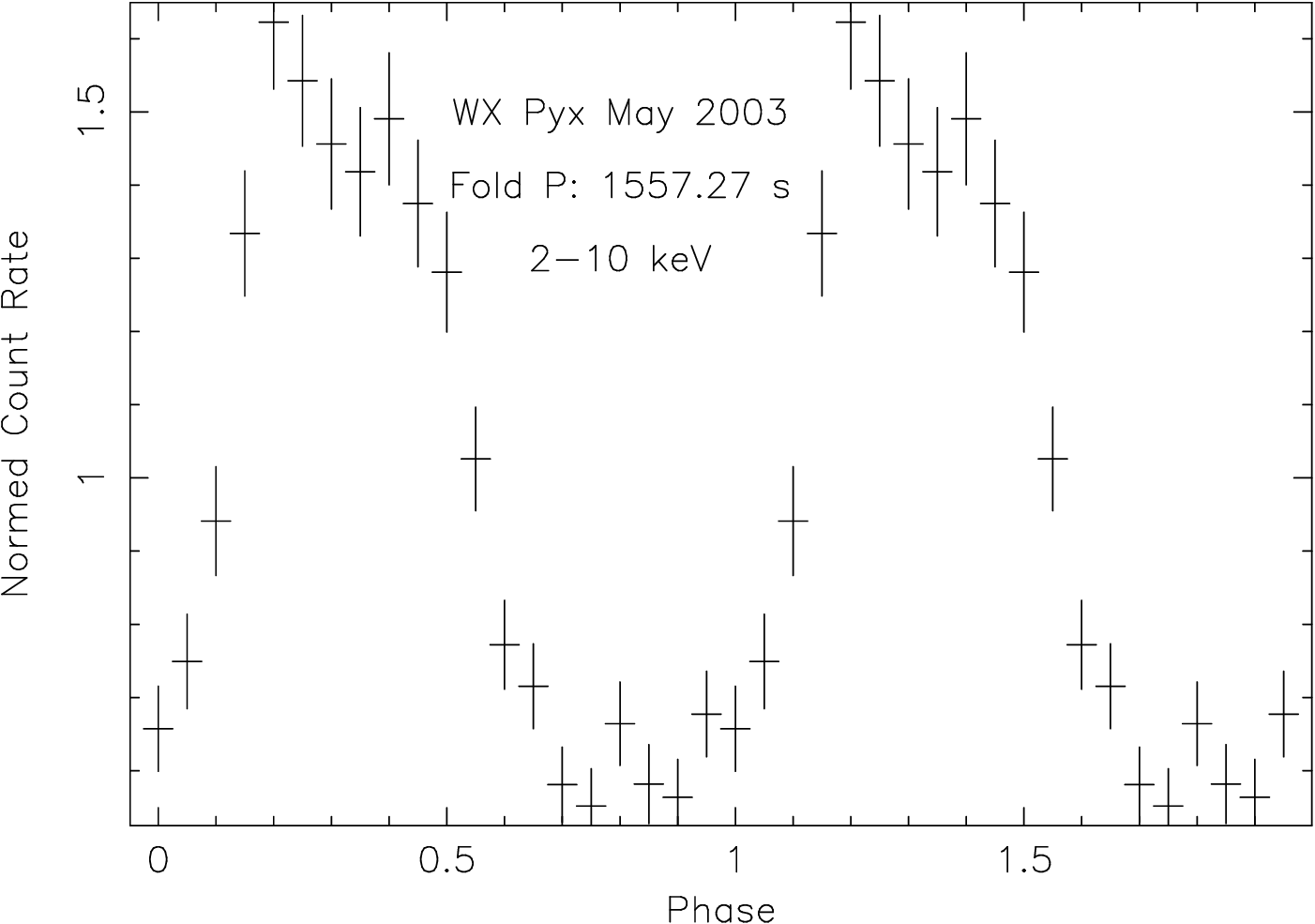}}}
\end{figure}

From the ${\omega}-{\Omega}$ and ${\omega}+{\Omega}$ values, the
orbital frequency is $\sim$5.0167${\times}$10$^{-5}$ s$^{-1}$
corresponding to an orbital period of $\sim$19933 sec $\sim$5.54 hour.
If we determine the spin frequency from the ${\omega}-{\Omega}$ and
${\omega}+{\Omega}$ values, that value differs from the fitted spin
frequency by $\sim$0.06\%.  Applying that difference to the orbital
period implies an estimated uncertainty in orbital period of
$\sim$15-20 sec.  That no power appears in the power spectrum at the
orbital period is not surprising given that the filtered observation
is at best $\sim$1.5 times the length of the orbital period.  The
counts in the soft and hard bands for each P$_{\rm spin}$ interval
were summed and a sinusoid fit to the resulting light curve to detect
or place a limit on orbital modulation.  The estimates of the orbital
modulation are 0.006$\pm$0.003 for the soft band and 0.014$\pm$0.008
for the hard band where the errors are 90\% for one parameter of
interest.  The modulation period was of the order of 20 ksec with
large errors owing to the rather few number of points in the orbital
curve ($\sim$19), but consistent with the derived candidate orbital
period.

Also note that a signal at $\sim$1230 sec (detected frequency =
8.121$\sim$10$^{-4}$ Hz) is also detected.  {\it XMM-Newton} does not
dither owing to the large EPIC pixels, so power at this frequency can
not be so attributed.  The difference between the detected frequency
and the spin frequency is $\sim$1.698$\times$10$^{-4}$ and does not
correspond to a low-order integer of the spin or orbital periods.

The soft and hard light curves from the May PN data were also folded
at the spin period.  Figure~\ref{foldLC} shows the results.  Both
bands exhibit sinusoid-like behavior.  The origin for each phase fold
used the \cite{odonoghue1996} definition for phase 0.0.  They defined
phase 0.0 to be the time of the maximum of the pulse.  As the time has
drifted by $\sim$0.35 in phase, the spin period is either known
insufficiently accurately or a small change in the period is present.
The April PN folded light curve is consistent in shape with the May
curve but with considerably less statistical precision.

\begin{table}
\centering
\caption{Simultaneous Energy Spectrum Fits}
\label{fluxes}
\begin{tabular}{l r r r r r r}
\hline \hline
       &            &     &   &    & \multicolumn{2}{c}{Flux} \cr
 Model & ${\chi}^2$/${\nu}$ & DoF & N$_{\rm H}$ & kT or ${\Gamma}$ & 0.5-2 & 2-10 \cr \hline
 Bremss & 0.81 & 9464 & $<$0.8 & 18$^{+24}_{-6}$ & 4.6 & 19 \cr
 Power Law & 0.60 & 9464 & $<$0.7 & 1.44$^{+0.06}_{-0.07}$ & 4.5 & 20 \cr \hline
\end{tabular}
Note:  N$_{\rm H}$ units = 10$^{20}$ cm$^{-2}$; unabsorbed
fluxes in units of 10$^{-13}$ erg s$^{-1}$ cm$^{-2}$.
\end{table}

\section{Spectral Fit}

\begin{table}
\centering
\caption{Epoch Fits to EPIC-PN Spectra}
\label{pnonly}
\begin{tabular}{l r r r r r r}
\hline \hline
       &            &     &    & Brems & \multicolumn{2}{c}{Flux} \cr 
 Month & ${\chi}^2$/${\nu}$ & DoF & N$_{\rm H}$ & kT & 0.5-2 & 2-10 \cr \hline
 Apr & 0.75 & 3949 & 1.4$^{+1.7}_{-1.2}$ & $<$70 & 0.6 & 2.6 \cr
May & 0.82 & 3949 & $<$0.8 & 23$^{+40}_{-8}$ & 4.4 & 13 \cr \hline
\end{tabular}
Note:  N$_{\rm H}$ units = 10$^{20}$ cm$^{-2}$; unabsorbed
fluxes in units of 10$^{-13}$ erg s$^{-1}$ cm$^{-2}$.
\end{table}

Energy spectra were extracted using apertures of radius $\sim$1$'$.5
centered on the source; from the calibration handbook, an aperture of
this size located $\sim$10$'$ offaxis will enclose $>$98\% of the
events including hard events near 9-10 keV.  As the object lies near
the chip edges, there is insufficient chip area for a surrounding
annulus.  The background was obtained from an aperture of identical
size positioned south of the source so that no overlap occurred
between the source and background apertures.  Redistribution matrices
and effective areas were built for each spectrum using the SAS version
6.0 routines {\tt rmfgen} and {\tt arfgen}. 

\begin{figure}
\centering
\caption{Simultaneous model fit to the EPIC-PN, -MOS1, -MOS2 spectra
from the May observation and the EPIC-PN spectrum from the April
observation. No credible evidence exists for any Fe emission, or any
other lines.  The May PN spectrum is uppermost of the stack; the April
PN spectrum lies among the MOS spectra.  The model shown does not
include a line component, so any line emission should appear in the
residuals.  }
\scalebox{0.3}{\rotatebox{-90}{\includegraphics{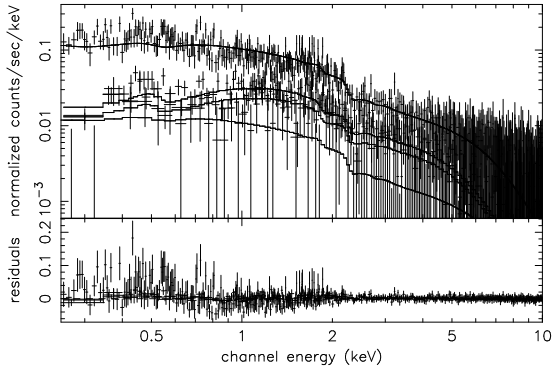}}}
\label{specfit}
\end{figure}

\begin{figure}
\centering
\caption{Contour plot for the bremsstrahlung temperature and column
density N$_{\rm H}$.  The contours correspond to the 1 ${\sigma}$,
90\%, and 99\% levels in ${\Delta}{\chi}^2$ for 2 parameters of interest.}
\scalebox{0.3}{\rotatebox{-90}{\includegraphics{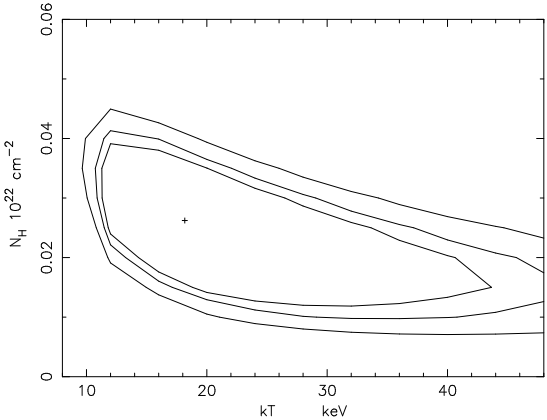}}}
\label{contbrem}
\end{figure}

A simultaneous fit was carried out on the four extracted spectra:
observation-1 PN; observation-2 MOS-1, MOS-2, and PN.  Channels less
than 0.25 keV and greater than 20.0 keV were ignored in the fit.  An
absorbed thermal bremsstrahlung and an absorbed power law model were
fit to the spectra; both provide reasonably good fits to the
continuum.  No evidence exists for an iron line, or any line
emission; a fit including a zero-width gaussian, to simulate an
unresolved line, leads to an upper limit on the presence of Fe line
emission of $\sim$300 eV.  Figure~\ref{specfit} shows the final fit
to the four spectra.  Note that the top-most histogram is the May PN
spectrum; the Apr PN spectrum lies about a factor of 10 lower.  The
contour plot for the bremsstrahlung temperature and column density
N$_{\rm H}$ is shown in Figure~\ref{contbrem}.

The mean fluxes in the 0.5-2 and 2-10 keV bands are
$\sim$4.6$\times$10$^{-13}$ and $\sim$2.3$\times$10$^{-13}$ erg
s$^{-1}$ cm$^{-2}$, as listed in Table~\ref{fluxes}.  The
corresponding luminosities are $\sim$5.5$\times$10$^{29}$ d$^2_{100}$
and $\sim$2.3$\times$10$^{30}$ d$^2_{100}$ erg s$^{-1}$, respectively,
using an adopted distance of 100 d$_{100}$ pc.  

Clearly visible is the difference in flux between the Apr and May PN
spectra, a factor of $\sim$10.  A fit to the spectra separately shows
a temperature broadly consistent between the two epochs (the Apr
observation only yields an upper limit), but the column density
differs.  The April observation produces a specific N$_{\rm H}$ value,
albeit with large error bars because of the low count rate.  The May
observation leads to an upper limit on the column density.  

For each epoch, upper limits may also be assigned to the
equivalent widths for potential line emission.  Adopting a gaussian
with zero width (to simulate an unresolved line), the 90\% upper
limits for the April PN spectrum are $\sim$100 eV at 1.7 keV (Si) and
$\sim$185 eV at 6.7 keV; the May PN values are $\sim$50 eV (Si) and
390 eV (Fe).  The high value in May corresponds to a higher count rate
in the hard band; an examination of the events in the 5.5-7.5 keV band
shows considerable noise, but no line emission.

\section{Discussion}

\cite{odonoghue1996} found a spin period of 1557.5$\pm$0.2 sec in
optical photometry from 1996.  The {\it XMM} data indicate a spin
period of 1557.3 sec and set a limit of $\sim$1.5$\times$10$^{-6}$ s
s$^{-1}$ on a period change over the intervening $\sim$8 years.  

The folded optical variation was nearly sine-like in the repeated
pulse shapes, but not exactly a sinusoid: the rise toward the peak
occurred slowly at first, then more rapidly.  This description mimics
the soft band {\it XMM} light curve (modulo the phase shift); if the
slope change identified by O'Donoghue et al. at phase $\sim$0.62
exactly matches that in the soft X-ray curve, then the phase shift is
$\sim$0.45, larger than but not inconsistent with the $\sim$0.35 shift
identified from the locations of the pulse peaks. 

The hard band light curve is more steep-sided, as if a hot region
that would generate a sinusoid light curve simply through geometric
projection, were buried such that the observer needed to be nearly
normal to it for it to be visible.  Simulated light curves (eg,
\cite{Kim1995}) show sinusoidal behavior at high energies and more
square-wave behavior at low energies.

The difference in flux by a factor of $\sim$10 in a few weeks is not
necessarily surprising given the high-low state behavior of IPs (eg,
GK Per, \cite{Hellier2004}).  The difference does not appear to
be confined to one portion of the spectrum as one might expect if, for
example, enhanced accretion led to a higher column density.  The flux
in both the soft and hard bands changed by similar factors.  By way of
an explanation, perhaps the {\it XMM} observation caught WX Pyx
emerging from a low state.

The strength of the spin component relative to the
${\omega}{\pm}{\Omega}$ components suggests disk accretion occurs in
WX Pyx given the weak orbital modulation \citep{Norton1996}.  That
supposition must be confirmed, however, with a longer observation as
the {\it XMM} observation only covers roughly 1.5 P$_{\rm orb}$.  A
weak orbital modulation likely indicates a low inclination, following
the results of, for example, \cite{Hellier1990}.  The strengths of the
sidebands are $\sim$0.3 and 0.2, respectively, for the
${\omega}-{\Omega}$ and ${\omega}+{\Omega}$ relative to the strength
of the spin period.  These values are well within the range of
behavior observed for disk-overflow IPs such as FO Aqr and imply that
a pure disk accretion interpretation is insufficient
\citep{Norton1996}.  Furthermore, given the factor of $\sim$10 change
in observed flux and if WX Pyx was emerging from a low state, then the
disk may have been recovering, thereby artificially weakening any
spin-orbit-coupled emission.  Long-term monitoring observations should
be useful to establish any high-low state changes and provide context
for the observations described here.

If the estimated orbital period of $\sim$19933 sec is confirmed, then
the P$_{\rm spin}$ / P$_{\rm orb}$ ratio is $\sim$0.08, placing WX Pyx
in the P$_{\rm spin}$-P$_{\rm orb}$ plane as a neighbor of the Cols,
TV and TX.  This is an intriguing trio of objects given their range of
observed behavior.  WX Pyx shows little iron emission and varies by a
factor of $\sim$10 in intensity over $\sim$month times.  TV Col
dsplays multiple periodicities and strong Fe emission
\citep{Rana2004}.  TX Col exhibits considerable variation in power
spectra components over times perhaps as short as a day
\citep{Schlegel2004,Schlegel2005}.  While the magnetic moment-P$_{\rm
spin}$ scenario described by \cite{Norton2004} is intellectually
unifying, there must be an additional parameter that dictates
different behavior from three IPs with nearly identical P$_{\rm spin}$
/ P$_{\rm orb}$ ratios.

\begin{acknowledgements}
The research of EMS was supported by contract number NAS8-39073 to
SAO. 
\end{acknowledgements}

\end{document}